\documentclass[journal]{IEEEtran}
\usepackage{graphicx}
\usepackage{epsf}
\usepackage{mathtools}
\usepackage{amssymb}
\usepackage{amsmath, amsfonts}
\usepackage{latexsym}
\usepackage{multirow}
\usepackage{multicol}
\usepackage[table]{xcolor}
\usepackage{color}

\usepackage{tabularx}
\usepackage{lipsum}

\usepackage{algorithm}
\usepackage{algorithmic}

\usepackage{mathrsfs}

\usepackage{fixltx2e}

\usepackage[mathscr]{euscript}

\usepackage{commath}
\usepackage{MnSymbol}

\usepackage{subcaption}

\usepackage{mathtools} 
\DeclarePairedDelimiterX\setc[2]{[}{]}{\,#1 \;\delimsize\vert\; #2\,}
\DeclarePairedDelimiterX\parth[2]{(}{)}{\,#1 \;\delimsize\vert\; #2\,}

\PassOptionsToPackage{hyphens}{url}
\usepackage[unicode=true, bookmarks=true, bookmarksnumbered=true, bookmarksopen=true, bookmarksopenlevel=1, breaklinks=false, pdfborder={0 0 0}, pdfborderstyle={}, backref=false, colorlinks=false]{hyperref}

\usepackage{theorem}
{
\theorembodyfont{\rmfamily}

}

\definecolor{orange}{RGB}{255,127,0}
\definecolor{blue}{RGB}{0,0,255}
\definecolor{red}{RGB}{255,0,0}
\definecolor{green}{RGB}{50,160,50}
\definecolor{grey}{RGB}{125,120,125}
\definecolor{purple}{RGB}{125,0,125}

\begin{document}
{
\title{{\fontsize{20}{2}\selectfont Is 30 MHz Enough for C-V2X?}}

\author
{
Dhruba Sunuwar, Seungmo Kim, \textit{Senior Member}, \textit{IEEE}, and Zachary Reyes

\vspace{-0.3 in}

\thanks{D. Sunuwar, S. Kim, and Z. Reyes are with the Department of Electrical and Computer Engineering, Georgia Southern University, Statesboro, GA, USA. The corresponding author is S. Kim who can be reached at seungmokim@georgiasouthern.edu. This work is supported by the Georgia Department of Transportation (GDOT) via grant RP21-08 and the National Science Foundation (NSF) via award ECCS-2138446.}
}

\maketitle
\begin{abstract}
Connected vehicles are no longer a futuristic dream coming out of a science fiction, but they are swiftly taking a bigger part of one's everyday life. One of the key technologies actualizing the connected vehicles is vehicle-to-everything communications (V2X). Nonetheless, the United States (U.S.) federal government decided to reallocate the spectrum band that used to be dedicated to V2X uses (namely, the ``5.9 GHz band'') and to leave only 40\% of the original chunk (i.e., 30 MHz of bandwidth) for V2X. It ignited concern of whether the 30-MHz spectrum suffices key V2X safety messages and the respective applications. This paper aims at addressing this issue with the particular focus on the New Radio (NR)-V2X Mode 1. We lay out an extensive study on the safety message types and their latency requirements. Then, we present our simulation results examining whether they can be supported in the 30-MHz spectrum setup.
\end{abstract}


\section{Introduction}\label{sec_intro}

\subsubsection{Background}
V2X technology allows vehicles to communicate with other vehicles, infrastructure, and vulnerable road users to enhance safety, thereby preventing traffic crashes, mitigating fatalities, alleviating congestion, and reducing the environmental impact of the transportation system \cite{usdot_17}. This capability gives V2X communications the central role in the constitution of intelligent transportation systems (ITS) for connected vehicle environments.

The full 75 MHz of the 5.9 GHz spectrum band (5.850-5.925 GHz) has long been reserved for intelligent transportation services such as V2X technologies. Nonetheless, the U.S. Federal Communications Commission (FCC) voted to allocate the lower 45 MHz (i.e., 5.850-5.895 GHz) for unlicensed operations to support high-throughput broadband applications (e.g., Wireless Fidelity, or Wi-Fi) \cite{nprm_19}. Moreover, the reform went further to dedicating the upper 30 MHz (i.e., 5.895-5.925 GHz) for cellular V2X (C-V2X) as the only technology facilitating ITS.

To this line, we deem it prudent to evaluate \textit{what is possible in a limited 30 MHz spectrum} to ensure that the ITS stakeholders can continue to develop and deploy these traffic safety applications.

\subsubsection{Related Work}
Obviously, C-V2X has been forming a massive body of literature. Nonetheless, only little attention was paid to the feasibility of C-V2X in the reduced 30 MHz spectrum for safety applications.

The end-to-end per-packet latency, defined as the time spent by a successful packet to travel from its source to final destination, is a classical networking metric. An advanced latency metric---namely, the inter-reception time (IRT)---was proposed, which measures the time length between successful packet deliveries \cite{irt_06}. However, we find the IRT to have limited applicability as it becomes efficient in broadcast-based safety applications only. Considering the variety of our target applications, this paper employs the classical latency as the main metric, as shall be detailed in Section \ref{sec_proposed}.

Now, in regard to the characterization of a V2X system, the literature introduced a wide variety of proposals. Several approaches were compounded into large bodies such as theoretical/mathematical approaches \cite{access19}-\cite{access20}, simulation-based \cite{gc18}\cite{faizan_iceic22}, and channel sounding-based \cite{nyu}. Even reinforcement learning has been applied as a means to characterize a highly dynamic vehicular network \cite{vtc20f}-\cite{nasim22}. The prior art certainly provides profound insights, yet is not directly conclusive whether the reduced 30-MHz bandwidth makes it feasible to operate C-V2X on realistic road and traffic scenarios.

The same limitation can be found in the current literature of V2X safety-critical applications \cite{ZoK22a}-\cite{milcom19}: the proposals lay out approaches to support such applications but leave it unaddressed what the influence will be after the C-V2X got deprived of 60\% of its bandwidth.

\subsubsection{Contributions}
This paper is a preliminary study as a starting point for discussions within the ITS literature in regard to operating C-V2X technologies in such an environment. As such, rather than final nor conclusive, this work should be regarded an initiative, igniting further tests and assessments on the impact of 30 MHz environment on the application deployment. Provided that, we extend the C-V2X literature on the following fronts:
\begin{itemize}
\item Pioneer to clarify the feasibility of safety-critical applications in the reduced 30 MHz spectrum setting
\item Develop a quantification framework for C-V2X performance in a comprehensive but easy manner to maneuver
\item Identify message types associated with safety-critical applications
\end{itemize}

\section{System Model}\label{sec_model}

\subsection{Spatial Setup}\label{sec_model_spatial}
Fig. \ref{fig_model} illustrates an urban environment setup that was used in this paper's simulations \cite{saej3161_1a}. A two-dimensional space $\mathbb{R}^2$ is supposed, which is defined by the dimensions of 520 m and 240 m for the north-south and the east-west axes, respectively.

Here is the summary of our spatial model shown in Fig. \ref{fig_model}. The RSUs are marked as green squares. The range of operation of each RSU is set to 150 m, indicated by a black circle around each RSU. There are two types of physical obstacles: trailer trucks (marked as black rectangles) and buildings (drawn as big gray squares).

We suppose \textit{two junctions} (rather than just one) as an effort to examine any possible interference between roadside units (RSUs) on the C-V2X performance as each junction is equipped with a RSU.

The connection from a RSU to a vehicle is shown by either a red or blue line: the red indicates a ``blocked'' connection whereas the blue means a ``connected'' link. The blockage can be caused by physical obstacles, viz., a building or a large trailer truck that are displayed by a large gray square and a black rectangle, respectively.

The distribution of the vehicles follows a \textit{homogeneous PPP} in $\mathbb{R}^2$. We define a general situation where a safety-critical application disseminates a message of its respective type over a C-V2X network. (See Section \ref{sec_proposed_message} for details on the message types.) Unlike vehicles, locations of RSUs are fixed each junction \cite{mdpi_rsu}.

By $\lambda$ and $\theta$, we denote the densities of vehicles and trucks per road segment, respectively. According to the densities $\lambda$ and $\theta$, the probability of signal being blocked varies, which, in turn, influences the end-to-end latency of a message. For instance, a large $\lambda$ and a large $\theta$ yield a higher level of competition for medium and an increased level of physical signal blockage, which therefore elevates the latency accordingly.

It is also noteworthy that each vehicle, upon reaching the end of a road, starts over from the opposite end of the same lane. This setup is to keep the total number of vehicles constant at all times, as a means to maintain the same level of competition for the medium at any given time and thus guarantee the accuracy for further stochastic analyses.

\begin{figure}[t]
\centering
\includegraphics[width = 0.8\linewidth]{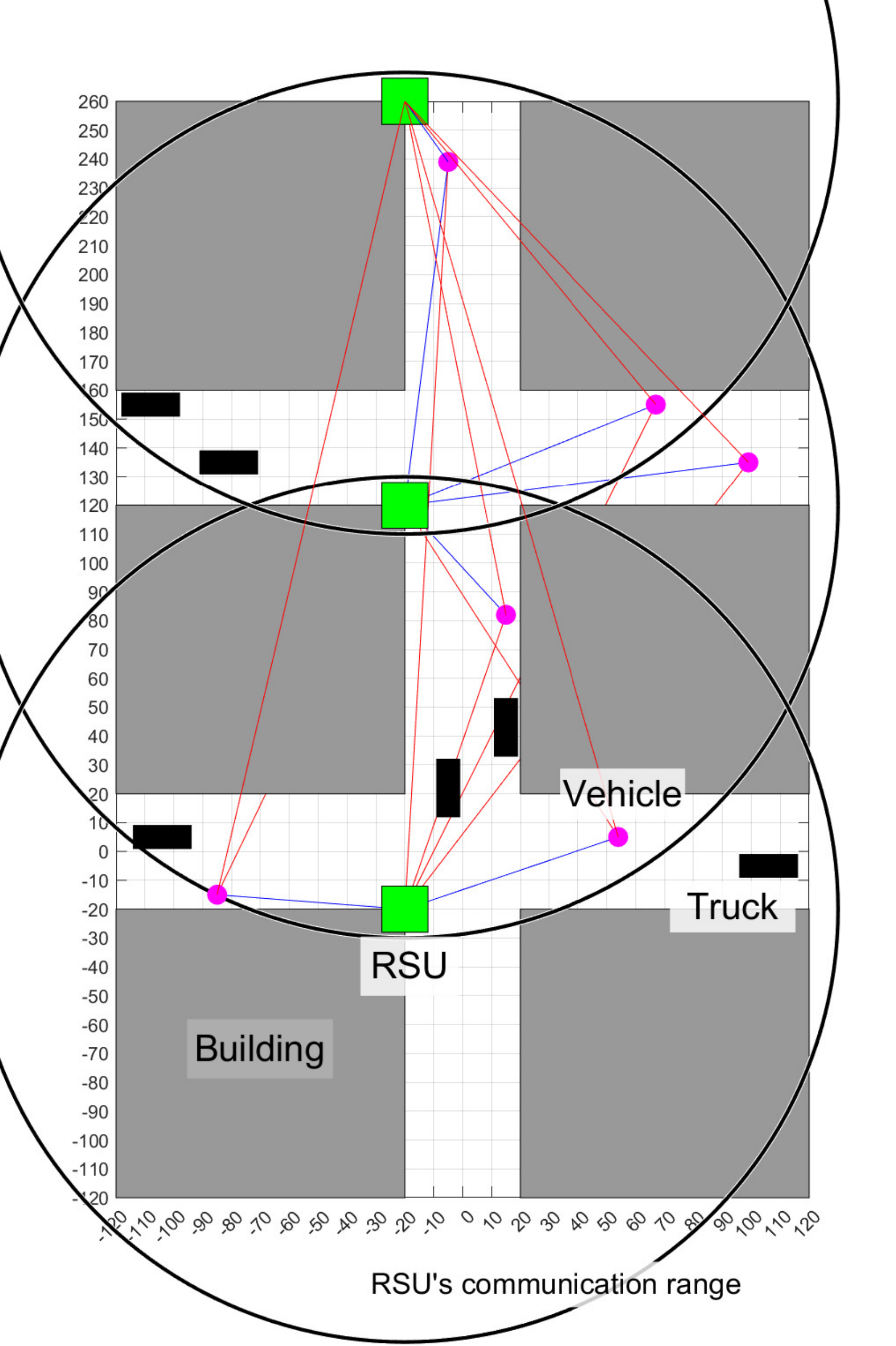}
\caption{Geometrical setup of the proposed simulator (with vehicle density of $\lambda = 6$ in the entire system space $\mathbb{R}^2$)}
\label{fig_model}
\vspace{-0.1 in}
\end{figure}

\subsection{Communications Parameters}\label{sec_model_communications}
This paper adopts the 3rd Generation Partnership Project (3GPP) Release 17 NR-V2X for the physical-layer (PHY) \cite{tr38901} and radio resource control (RRC) functions \cite{ts36101}. We assume \textit{Mode 1} where a NR base station (also known as ``Next Generation Node B'' or ``gNB'') schedules sidelink resources to be used by the user equipment (UE) (i.e., vehicle) for sidelink transmissions. Nonetheless, we claim that the versatility of our simulation framework can easily be extended to accommodate Mode 2 as well.

To elaborate on the sidelink of NR-V2X, our simulation implements all the key channels \cite{ts38201} including: the Physical Sidelink Broadcast Channel (PSBCH) for sending broadcast information (i.e., synchronization of the sidelink); the PSCCH for sending control information; the PSSCH for sending control; data and Channel State Information (CSI) in case of unicast; and the Physical Sidelink Feedback Channel (PSFCH) for sending HARQ feedback in case of unicast and groupcast modes.

We suppose that all the vehicles distributed in $\mathbb{R}^2$ have the same ranges of carrier sensing and communication. The NR-V2X supports \{10, 15, 20, 25, 50, 75, 100\} resource blocks (RBs) per subchannel \cite{ts38202}. As shall be elaborated in Section \ref{sec_results}, this paper supposes 50 RBs per subchannel, which matches our assumption of 10 MHz per channel.

Our simulation also features a very high level of precision in implementing the spatial environment. Since a city road environment is considered for the simulation as shown in Fig. \ref{fig_model}, the Urban Micro (UMi)-Street Canyon path loss model \cite{tr38901} is implemented. However, we reiterate that our simulation can easily accommodate other path loss models defined in the standard \cite{tr38901}.

\begin{table*}[t]
\centering
\caption{Mapping of message types to traffic priority levels}
\begin{tabular}{|c||c|c|c|c|c|c|c|}
\hline
\rowcolor{gray!20} Service Type & \multicolumn{4}{|c|}{Safety Services} & \multicolumn{3}{|c|}{Mobility Services}\\\hline
\rowcolor{gray!20} Traffic Direction & \multicolumn{2}{|c|}{V2V} & \multicolumn{2}{|c|}{V2I-I2V} & \multicolumn{3}{|c|}{V2I-I2V}\\\hline
\cellcolor{gray!20} Traffic Families & Critical V2V & Essential V2V & Critical V2I-I2V & Essential V2I-I2V & Transactional & Low-priority & Background\\\hline
\cellcolor{gray!20} Minimum PPPP & 2 & 5 & 3 & 5 & 6 & 6 & 8\\\hline
\cellcolor{gray!20} Minimum PDB & 20 msec & 100 msec & 100 msec & 100 msec & 100 msec & 100 msec & 100 msec\\\hline
\cellcolor{gray!20} Example Messages & BSM, EVA & BSM & RSM, MAP & SPaT, RTCM & SSM, SRM & TIM, RWM & TCP, UDP\\\hline
\end{tabular}
\label{table_message_types}
\end{table*}

\section{Proposed NR-V2X Performance Evaluation Framework}\label{sec_proposed}
\subsection{Message Types}\label{sec_proposed_message}
Table \ref{table_message_types} categorizes several representative message types (i.e., as the last row) according to the ``traffic families'' (i.e., as the 3rd row). We particularly highlight that the ongoing SAE J3161 standardization activity \cite{saej3161_1a} is primarily based on the end-to-end latency, namely, the packet delay budget (PDB). Discussion on the metric selection shall be revisited in Section \ref{sec_proposed_metric}.

We assign different ProSe per-packet priorities (PPPP) \cite{ts23303} based on the importance of a message type. This proposition is to further extension to optimization of C-V2X via assigning different communication profiles (viz., number of subchannels, modulation and coding scheme (MCS), number of retransmissions, etc.) for the packets based on packet size, velocity, and channel busy ratio (CBR).

Here is elaboration of Table \ref{table_message_types} \cite{saej2735}\cite{usdot_cv273}: basic safety (BSM), emergency vehicle alert (EVA), road safety (RSM), map data (MAP), signal phase and timing (SPaT), Radio Technology Commission for Maritime Services corrections (RTCM), signal status (SSM), signal requeset (SRM), traveler information (TIM), and road weather (RWM), as well as even transport-layer protocols such as transmission control (TCP) and user datagram (UDP). These types of messages support a broad set of vehicle-to-vehicle (V2V) and vehicle-to-infrastructure (V2I) applications, e.g., forward collision warning, pre-crash sensing, emergency vehicle warning and signal preemption, and infrastructure-to-vehicle warning messages.

As found from the ``V2V'' column of Table \ref{table_message_types}, some applications operate based on the same message types, allowing numerous applications to be operated without requiring additional spectrum. However, different applications using the same message types can have vastly different spectrum needs due to differing message sizes and frequency of message transmission, so there are scenarios in which some applications using the same message types could and could not be deployed. Additionally, available spectrum will be dependent in part on the number of vehicles within communication range and the types of applications operating in a given area. Because of this, it will likely be necessary to establish a scheme that prioritizes safety-critical applications while underrating non-safety-critical applications in such situations.


\begin{figure}
\centering
\includegraphics[width = 0.7\linewidth]{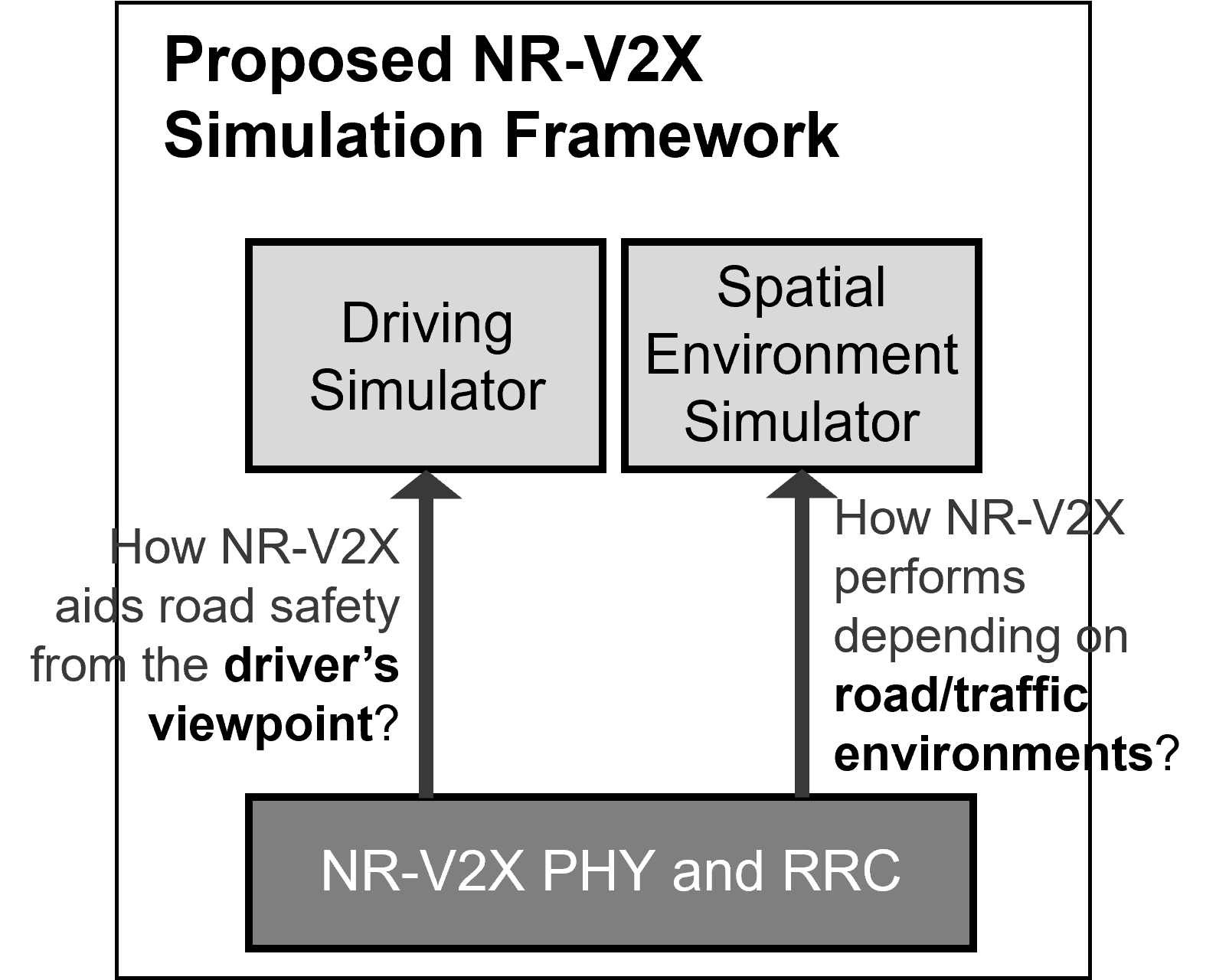}
\caption{Structure of the proposed simulator}
\label{fig_simulator}
\end{figure}

\begin{figure*}
\centering
\begin{subfigure}[b]{0.32\linewidth}
\centering
\includegraphics[width = \linewidth]{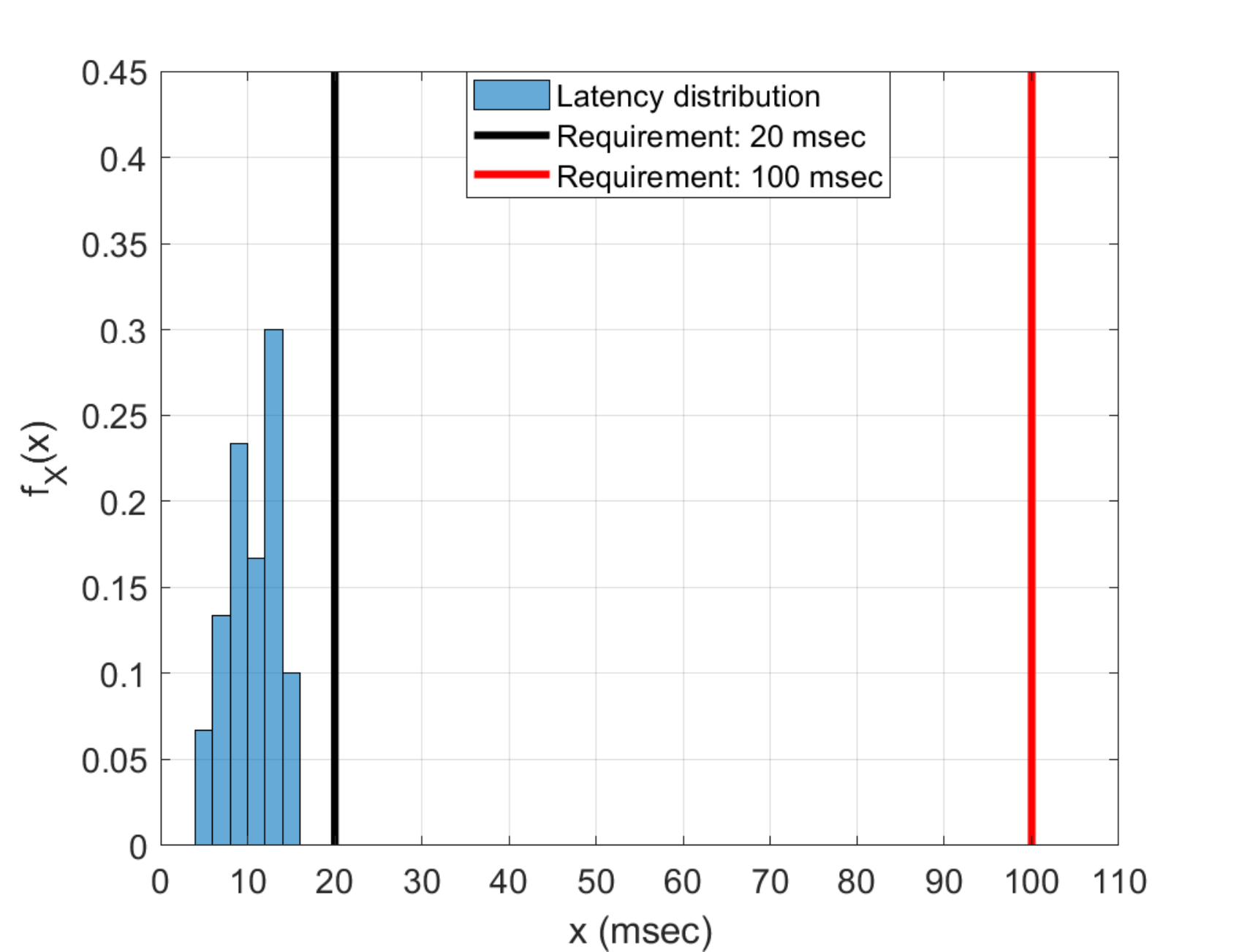}
\caption{1 RSU and 5 vehicles/lane}
\label{fig_latency_rsu1_lambda5}
\end{subfigure}
\begin{subfigure}[b]{0.32\linewidth}
\centering
\includegraphics[width = \linewidth]{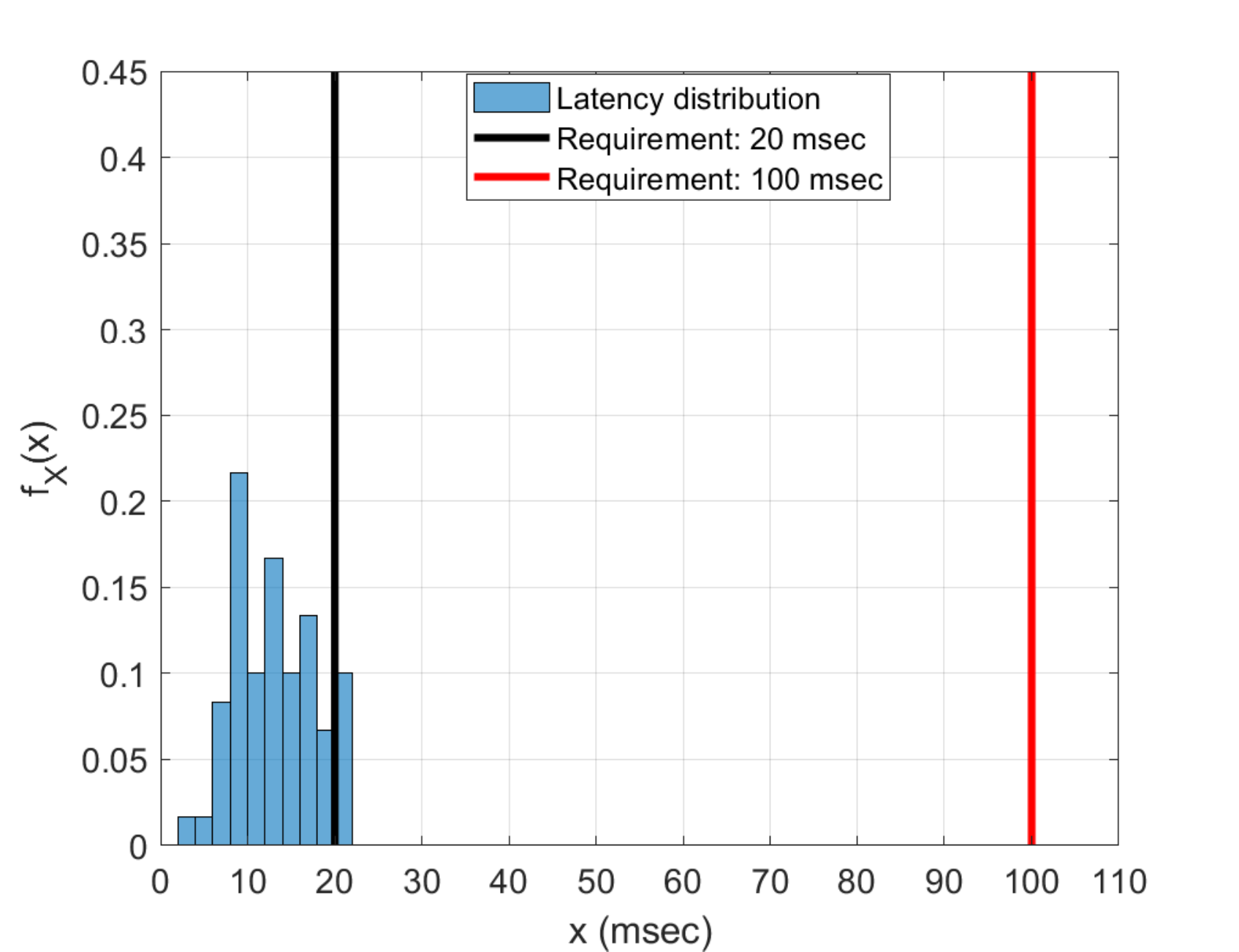}
\caption{1 RSU and 10 vehicles/lane}
\label{fig_latency_rsu1_lambda10}
\end{subfigure}
\begin{subfigure}[b]{0.32\linewidth}
\centering
\includegraphics[width = \linewidth]{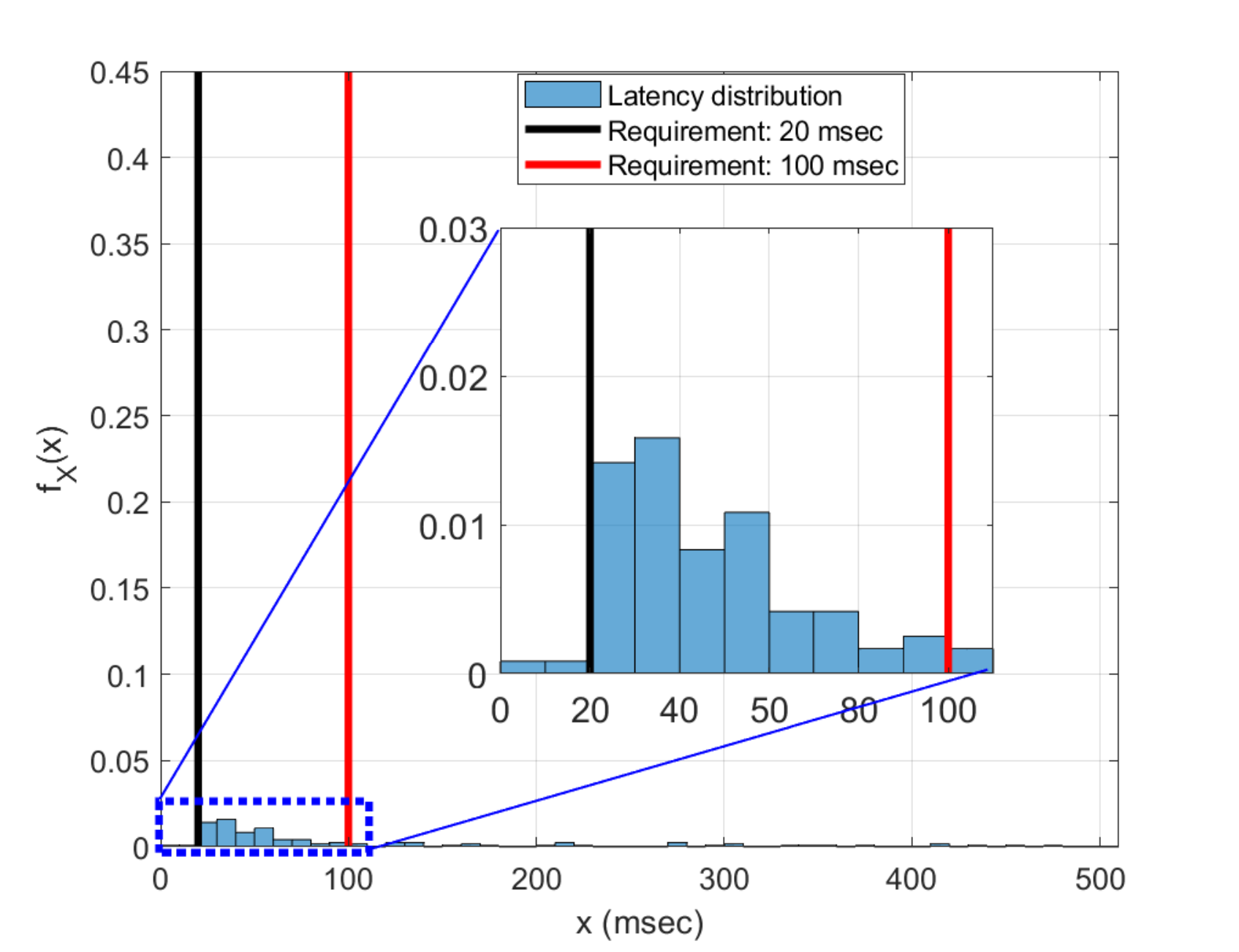}
\caption{1 RSU and 20 vehicles/lane}
\label{fig_latency_rsu1_lambda20}
\end{subfigure}
\begin{subfigure}[b]{0.32\linewidth}
\centering
\includegraphics[width = \linewidth]{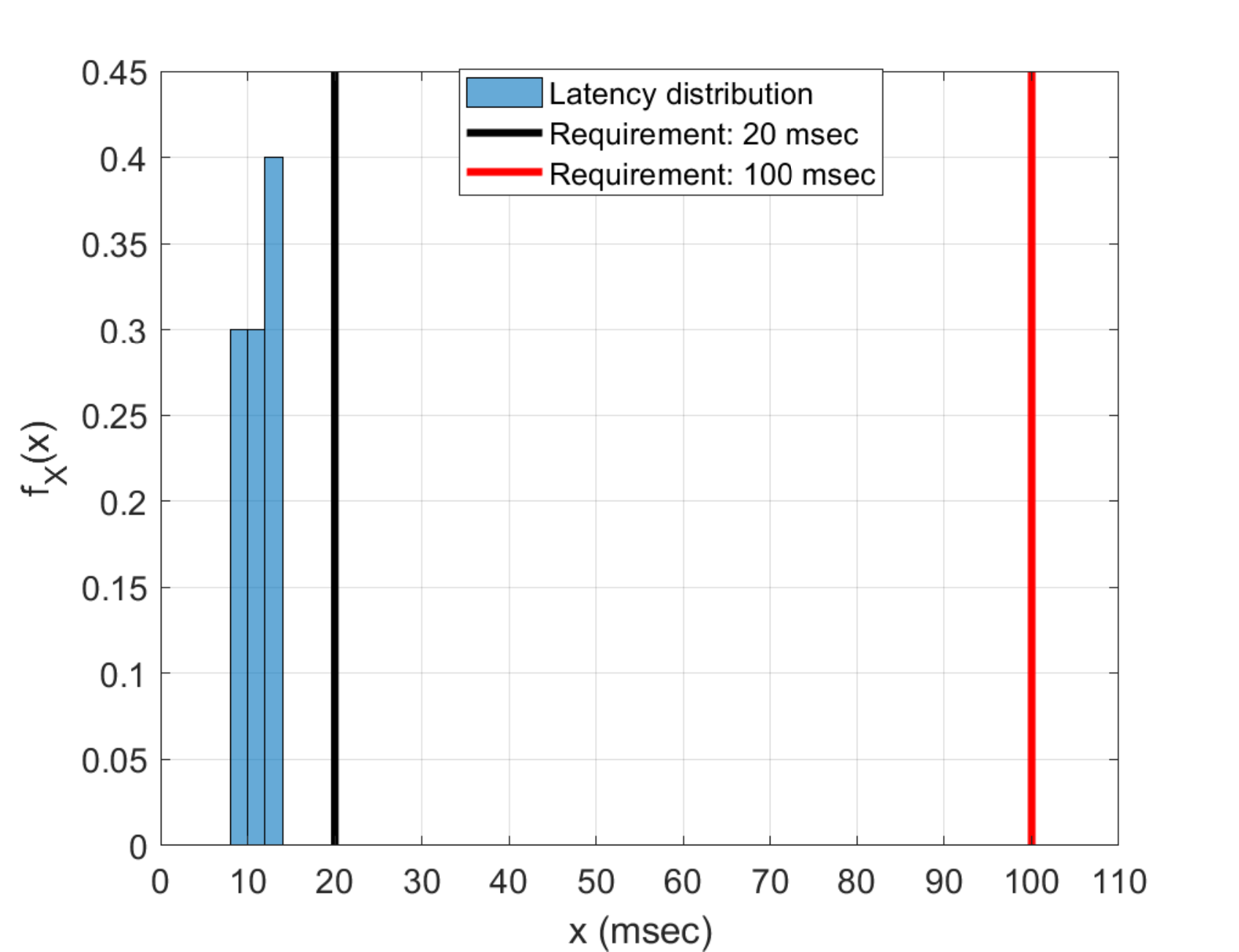}
\caption{2 RSUs and 5 vehicles/lane}
\label{fig_latency_rsu2_lambda5}
\end{subfigure}
\begin{subfigure}[b]{0.32\linewidth}
\centering
\includegraphics[width = \linewidth]{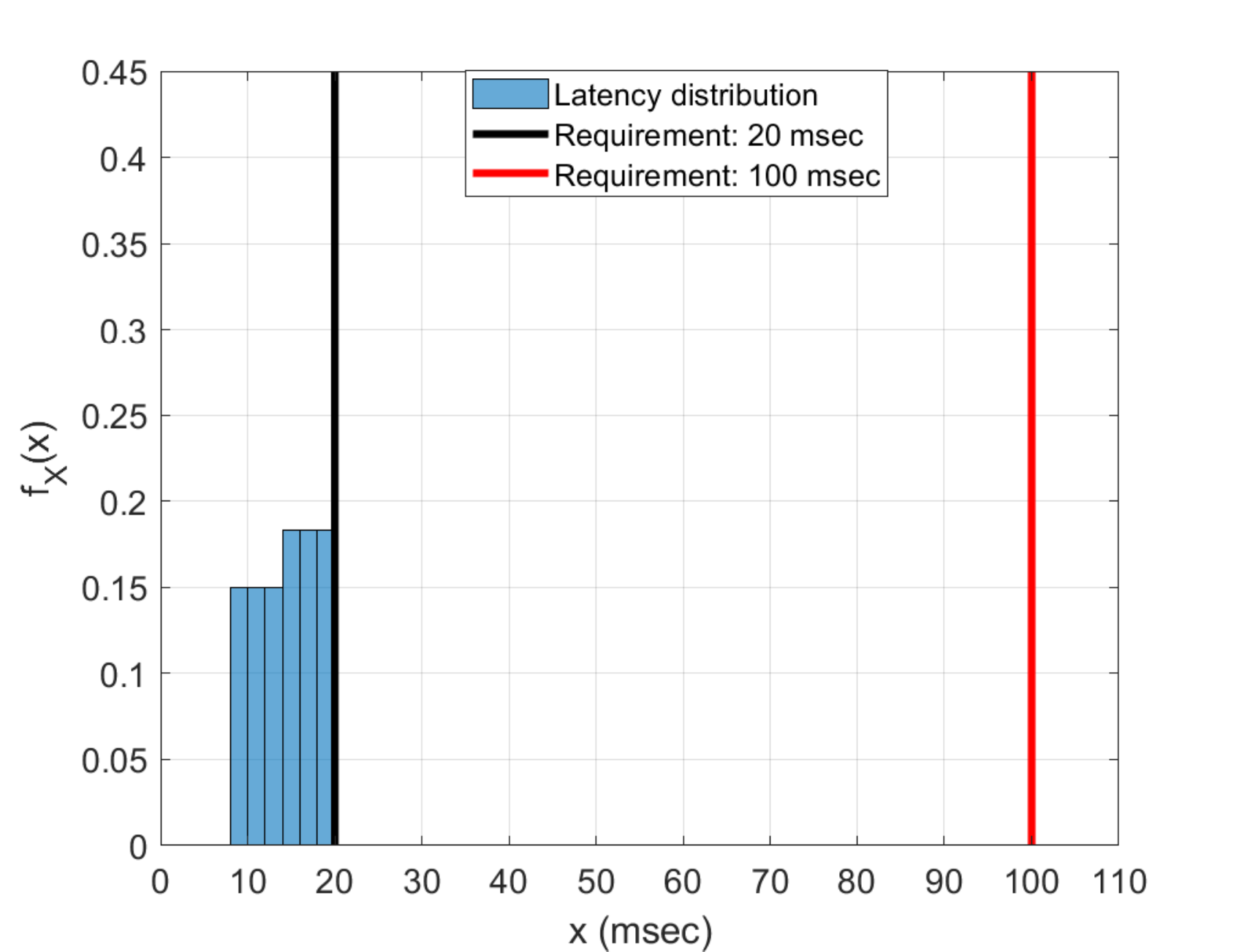}
\caption{2 RSUs and 10 vehicles/lane}
\label{fig_latency_rsu2_lambda10}
\end{subfigure}
\begin{subfigure}[b]{0.32\linewidth}
\centering
\includegraphics[width = \linewidth]{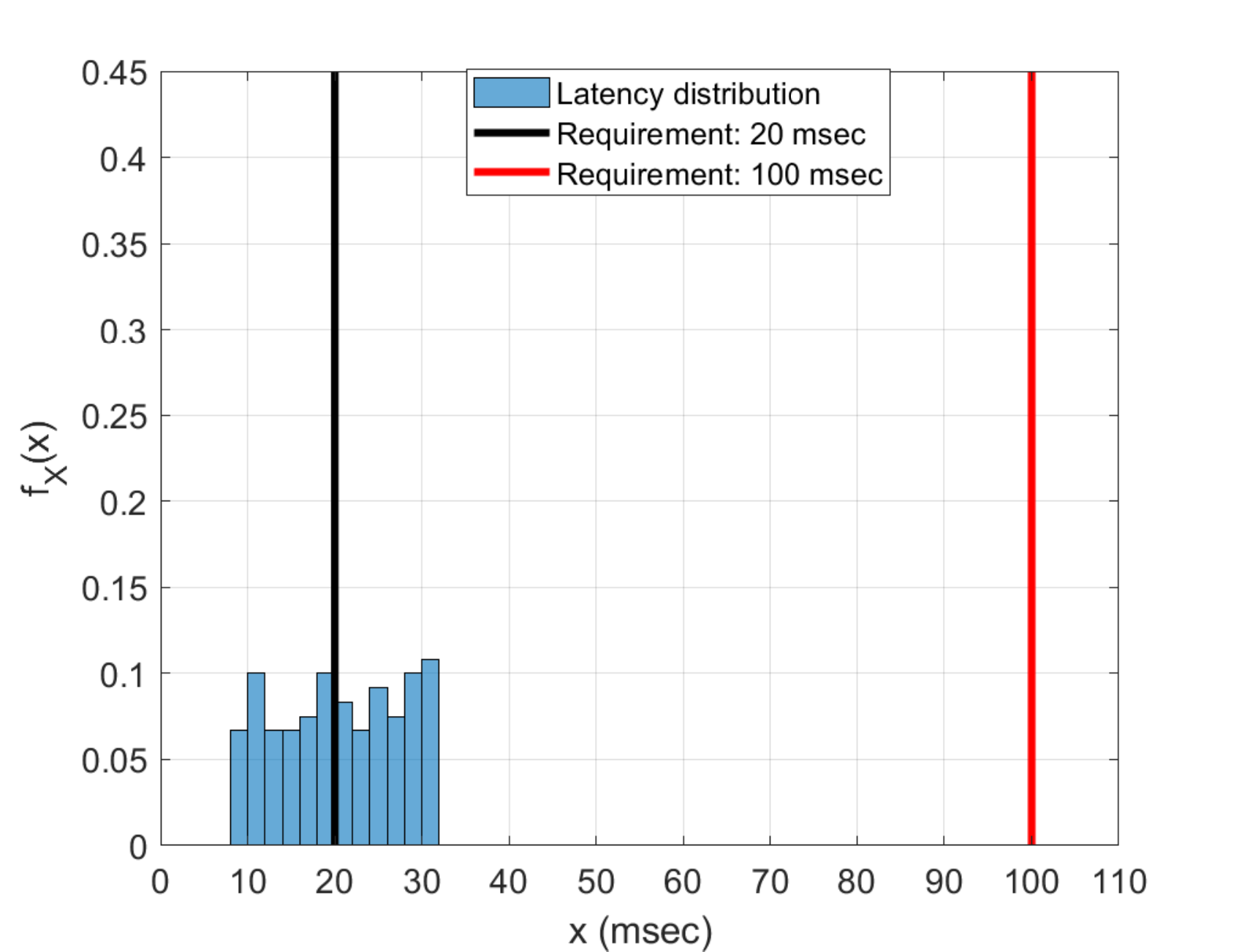}
\caption{2 RSUs and 20 vehicles/lane}
\label{fig_latency_rsu2_lambda20}
\end{subfigure}
\begin{subfigure}[b]{0.32\linewidth}
\centering
\includegraphics[width = \linewidth]{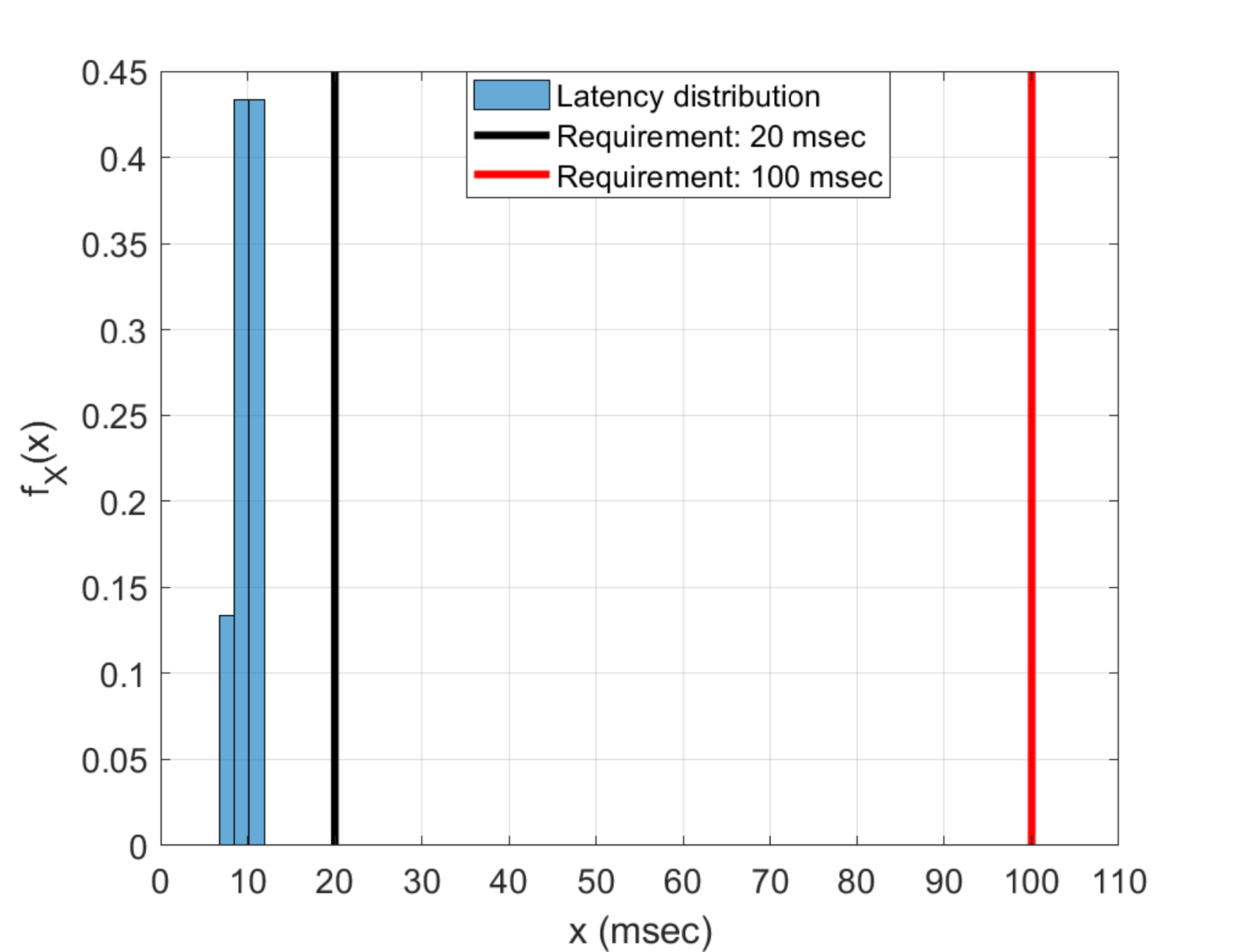}
\caption{3 RSUs and 5 vehicles/lane}
\label{fig_latency_rsu3_lambda5}
\end{subfigure}
\begin{subfigure}[b]{0.32\linewidth}
\centering
\includegraphics[width = \linewidth]{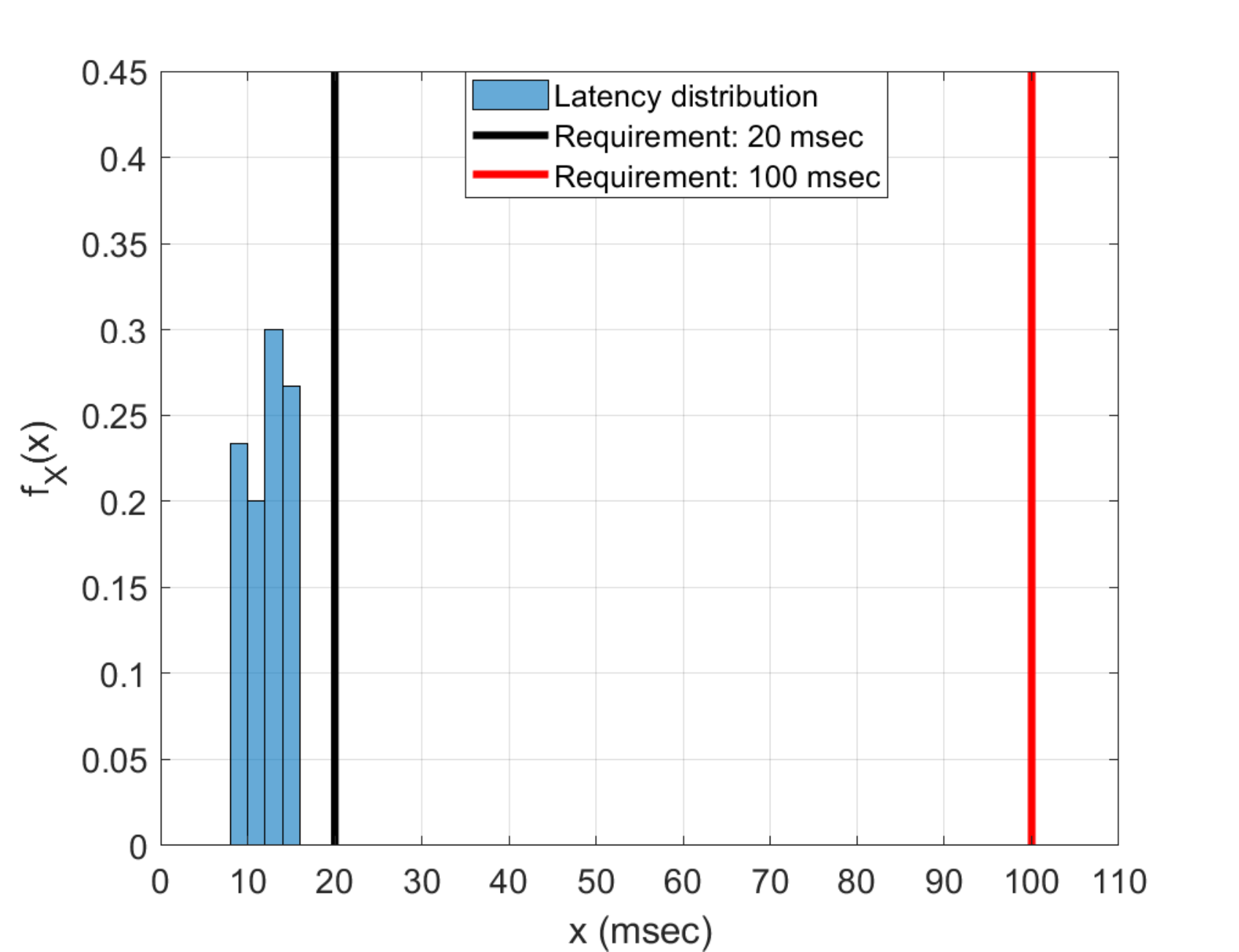}
\caption{3 RSUs and 10 vehicles/lane}
\label{fig_latency_rsu3_lambda10}
\end{subfigure}
\begin{subfigure}[b]{0.32\linewidth}
\centering
\includegraphics[width = \linewidth]{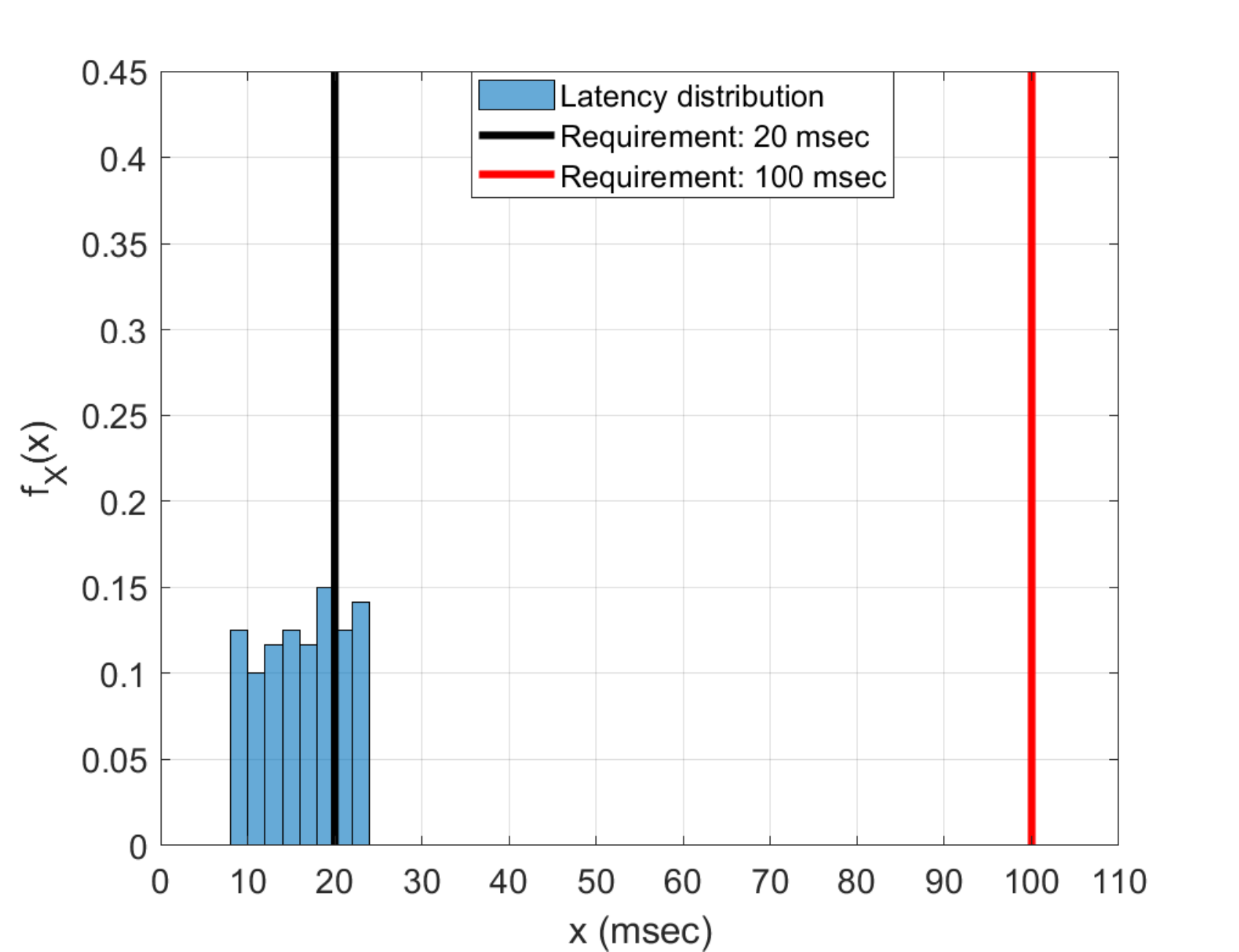}
\caption{3 RSUs and 20 vehicles/lane}
\label{fig_latency_rsu3_lambda20}
\end{subfigure}
\caption{Distribution of RSU-to-vehicle latency according to number of RSUs (compared column-wise) and traffic density (compared row-wise) compared to latency requirements of \{20,100\} msec as \{black,red\} vertical lines}
\label{fig_latency}
\end{figure*}

\subsection{Simulator Development}
This proposed research features \textit{integration} of the NR-V2X PHY and RRC simulator with two other major simulators: namely, the geometrical simulator and the driving simulator. Fig. \ref{fig_simulator} illustrates how this integration is achieved.

First of all, the NR-V2X PHY and RRC simulator implements the communications functions that were mostly explained earlier in Section \ref{sec_model_communications}. As such, it effectuates the sidelink communications among the vehicles and RSUs, following major TRs and TSs \cite{ts38202}-\cite{ts38201} of the 3GPP Release 17 standard. This NR-V2X simulator forms the basis for two other major components of the proposed simulation framework.

On top of the NR-V2X simulator, the spatial environment simulator facilitates an urban environment shown in Section \ref{sec_model_spatial}. As Fig. \ref{fig_simulator} depicts, the existence of this spatial environment component adds the context of \textit{the NR-V2X performance in different road/traffic settings}. We emphasize that this component will be strengthened by adding a wider variety of road environments and traffic scenarios.

Now, we highlight that our simulation framework features the driving simulator that actually puts a human user into a 1st-person driving setup. That way, the user can have \textit{live experience of connected vehicles}: the experience can actualize the user on how car-to-car connections can promote the safety in various traffic scenarios and road conditions. As shall be elaborated in Section \ref{sec_results_driving}, this driving simulator will also play the role of adding realistic contexts to V2X simulations, which clearly highlights the unique contribution of this research.

We combine all of those so the user can not only (i) deploy vehicles, RSUs, and obstacles but (ii) promptly quantify the C-V2X performance out of the scenario.

\subsection{Metrics}\label{sec_proposed_metric}
What it takes to call the V2X performance ``enough''? It is critical to address to this question in order to address what this paper's title asks: \textit{is 30 MHz enough for NR-V2X Mode 1?} We stress that this paper adopts the \textit{end-to-end latency} as the primary performance metric measuring a C-V2X system.

The end-to-end latency is the length of maximum allowed time between the generation of a message at the transmitter's application and the reception of the message at the receiver's application \cite{surveys_22}. As this paper focuses on Mode 1 of the NR-V2X, we implement the the latency as the length of time taken from the generation of a message at an application (of those listed in Table \ref{table_message_types}) at a RSU to the reception of the message by a vehicle.

Here is the justification of ``why'' the latency is chosen as the key performance metric in this paper over other metrics. First and foremost, the 3GPP 5G Service Requirement also identifies the end-to-end latency as one of the most critical performance indicators \cite{ts22186}, based on which other requirement factors are defined. Not only that, the ongoing SAE J3161 standardization activity \cite{saej3161_1a} is almost solely based on the latency, i.e., PDB.

However, in addition to the delay, we also reiterate that this paper features the integration with driving simulator. We measure the \textit{near crash rate} after a sufficiently large number of driving simulations on human subjects. Notice that a near-crash is defined as any circumstance that requires a rapid, evasive maneuver by the subject vehicle, or any other vehicle, pedestrian, cyclist, or animal to avoid a crash \cite{nearcrash_06}. A rapid, evasive maneuver is defined as a steering, braking, accelerating, or any combination of control inputs that approaches the limits of the vehicle capabilities. This helps add contexts on how the improved performance of C-V2X can \textit{actually affect the road safety}. More details on the simulation scenario follows in Section \ref{sec_results_driving}.

\begin{table}[t]
\caption{Key parameters for simulation}
\centering
\begin{tabular}{|l|l|}
\hline
\cellcolor{gray!20}Parameter & \cellcolor{gray!20}Value\\
\hline\hline
Inter-broadcast interval & 100 msec \cite{ts36101}\\
Bandwidth per channel & 10 MHz \cite{nprm_19}\\
Number of RBs per subchannel & 50 \cite{ts38202}\\
Payload length & 40 bytes \cite{saej2735}\\
Vehicle density & \{5,10,20\} vehicles/lane\\
\hline
\end{tabular}
\label{table_parameters}
\end{table}

\section{Numerical Results}\label{sec_results}

\subsection{NR-V2X RSU-to-Vehicle Latency}
Table \ref{table_parameters} summarizes the key parameters that were used in our NR-V2X simulation. Notice from the table that we assume 50 RBs per subchannel, which occupies 180 kHz/RB $\times$ 50 RBs/subchannel $\approx$ 9 MHz and thus takes up most of an entire 10-MHz channel considering 1.25 MHz of a guard band \cite{ts22186}. The vehicle density is another noteworthy parameter: $\lambda =\{5,10,20\}$ vehicles per lane equal $\{30,60,120\}$ vehicles in $\mathbb{R}^2$, which in turn indicate $\{48,24,12\}$ m of minimum and $\{76,38,19\}$ m of maximum inter-vehicle distance. As such, we intend that the $\lambda =\{5,10,20\}$ vehicles per lane represent the \{low, medium, high\} vehicle density, respectively.

Fig. \ref{fig_latency} demonstrates the result of this simulation. Each subfigure presents $f_{X}(x)$, the probability density function (pdf) of the RSU-to-vehicle latency $x$ in miliseconds (msec). The pdf is compared to the latency requirements of \{20,100\} msec as (\{black,red\} vertical lines}) that have earlier been discussed in Table \ref{table_message_types}. Via the comparison, Fig. \ref{fig_latency} displays very clearly \textit{how much proportion of vehicles generated on the $\mathbb{R}^2$ is able to support which message types and applications}.

Horizontal comparison in a single row indicates that a higher vehicle density yields a higher latency and thus a higher chance of exceeding the 20-msec latency requirement. Vertical comparison in a single column implies that a larger number of RSUs (each taking a full 10-MHz channel) gives a lower latency and thus a lower probability of exceeding the latency requirement.

We deem it a safe statement if one finds that even in the 30-MHz setting, the applications requiring the latency of 100 msec can be supported in most scenarios. The only exception from this study was the ``1 RSU and 20 vehicles/lane'' case; Fig. \ref{fig_latency_rsu1_lambda20} implies that such high vehicle density may cause some portions of vehicles to be served at a higher latency than the 100-msec requirement. Henceforth, it translates that in the 30-MHz spectrum setting, while almost all other message types defined in Table \ref{table_message_types} can be supported, the ``V2V critical BSM'' may not function with only one RSU in a high traffic density scenario.

\subsection{Verification via Driving Simulation}\label{sec_results_driving}
We are developing various driving simulation scenarios for testing the message types identified in Table \ref{table_message_types}. As has been mentioned in Fig. \ref{fig_simulator}, such \textit{integration} of a driving simulator and a NR-V2X simulator forms key merit of this research.

A usual setup for our driving simulator is as follows. A human subject is located in front of a driving simulator that we built. There are 4 monitors, a wheel, and pedals, emulating the windshield, the directional and speed maneuvers, respectively.

\begin{figure}
\centering
\includegraphics[width = .85\linewidth]{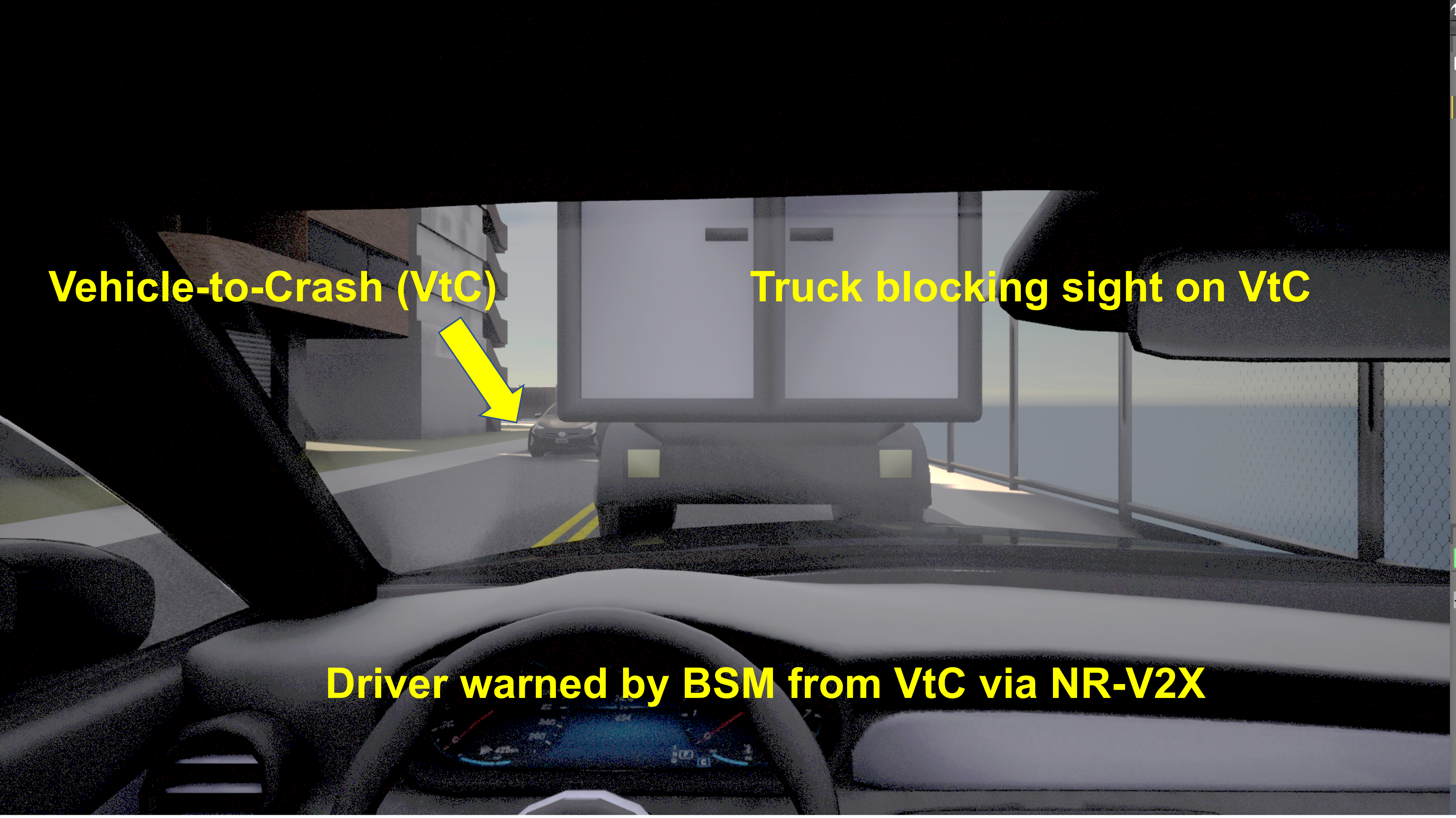}
\caption{``Do not pass'' warning scenario for V2V BSM test}
\label{fig_donotpass}
\end{figure}

Fig. \ref{fig_donotpass} illustrates the proposed driving simulation setup for the verification. The driven vehicle (the ``vehicle-of-interest'' or ``VoI'' hereafter) is put into the following scenario where continued exchange of V2V BSM via NR-V2X can improve the safety. The VoI is put onto a 2-lane road: i.e., a single lane per direction. The driver cares to pass the large trailer truck in front of the VoI, which blocks the driver's sight of the other-direction lane. We design the simulation that because of the sight blockage, an attempt to pass the truck causes a \textit{near crash} with another vehicle approaching from the other lane (the ``vehicle-to-crash'' or ``VtC''). The assumption here (which is very realistic) is that if the VoI and VtC have successfully exchanged BSMs, the near crash can be avoided by the driver being able to refrain from the pass with the VtC approaching.

\section{Conclusions}\label{sec_conclusions}
This paper proposed a framework for evaluating the performance of a NR-V2X system. The research was particularly motivated from the U.S. federal government's recent decision of leaving only 30 MHz of spectrum for V2X. As such, this research identified key V2X safety messages and the respective applications, and examined whether they can be still supported with the reduced bandwidth. In an urban Mode 1 setting, most of the safety-critical applications appeared to satisfy their latency requirements. Our holistic simulation framework integrating the NR-V2X PHY/RRC, spatial environment, and driving simulators strengthened the validity of the results.

As future work, we plan on further improvement of this simulator suite such that it accommodates a wider variety of NR-V2X functionalities, road conditions, and traffic scenarios, e.g., Mode 2, suburban highway, etc.

\section*{Acknowledgement}
We acknowledge the Intelligent Transportation Society of America (ITSA) and the delegates from their member organizations including Qualcomm, Cisco, etc. for valuable feedback via continued discussions on the C-V2X message types and traffic families that were presented in Section \ref{sec_proposed_message}.


\end{document}